\documentclass[aps,prd,twocolumn,superscriptaddress]{revtex4}
\usepackage{graphicx}
\usepackage{subfigure}
\usepackage{float}
\usepackage{amsmath}
\usepackage{amsfonts}
\usepackage{color}
\usepackage{bm}
\usepackage{bbm}
\usepackage{txfonts}
\usepackage{array}
\usepackage{amssymb}
\usepackage{braket}
\usepackage{multirow}
\usepackage{appendix}
\usepackage{tikz,xcolor,hyperref}
\definecolor{lime}{HTML}{A6CE39}
\DeclareRobustCommand{\orcidicon}{%
    \begin{tikzpicture}
    \draw[lime, fill=lime] (0,0)
    circle [radius=0.16]
    node[white] {{\fontfamily{qag}\selectfont \tiny ID}};    \draw[white, fill=white] (-0.0625,0.095)
    circle [radius=0.007];    \end{tikzpicture}
    \hspace{-2mm}}
\foreach \x in {A, ..., Z}{%
    \expandafter\xdef\csname orcid\x\endcsname{\noexpand\href{https://orcid.org/\csname orcidauthor\x\endcsname}{\noexpand\orcidicon}}
    }

\begin{document}
\title{Resonant tunneling of deuteron-triton fusion in strong high-frequency electromagnetic fields}
\author{Binbing Wu}
\affiliation{Graduate School, China Academy of Engineering Physics, Beijing 100193, China}
\author{Hao Duan}
\affiliation{Institute of Applied Physics and Computational Mathematics, Beijing 100088, China}
\affiliation{Laboratory of Computational Physics, Institute of Applied Physics and Computational Mathematics, Beijing 100088, China}
\author{Jie Liu\orcidA{} \footnote{Corresponding author: jliu@gscaep.ac.cn}}
\affiliation{Graduate School, China Academy of Engineering Physics, Beijing 100193, China}
\date{\today}
\begin{abstract}
We investigate deuteron-triton (DT) fusion in the presence of linearly polarized strong electromagnetic fields in high-frequency limit, in which  a complex spherical square-well potential  is exploited to describe the nuclear potential. Within the framework of the Kramers-Henneberger (KH) transformation, we have  calculated the total and  angular differential fusion cross sections  by investigating the asymptotical  phase shifts of the  Coulomb wavefunctions. With introducing a dimensionless quantity of $n_d$ representing the ratio of the particle quiver oscillation amplitude to the radius of nuclear potential, we find that, even though the tunneling probability of passing through the Coulomb repulsive potential  keeps almost identical to that in the absence of electromagnetic fields, the  peak of  total fusion sections shows an apparent shift from the well known value of 110 keV  to 78 keV for $n_d=0.01$. The angular differential cross sections also show some resonance peaks that shift from zero inclination  angle to $\pi/2$ with increasing the parameter $n_d$. The corresponding astrophysical $S$-factors are found to be enhanced by several times in amplitudes. With the help of  Wentzel-Kramers-Brillouin (WKB) approximate wavefunctions, the  shape-resonance tunneling mechanism of the above findings are uncovered and some implications are discussed.
\end{abstract}
\maketitle

\section{Introduction}
 With the rapid advance of intense laser technologies, particularly chirped pulse amplification technique, the laser intensities can reach up to
 $10^{22}{\rm W/cm^2}$ nowadays, and  another one or two orders of magnitude increase are expected  in the near future \cite{http:eli}. These intense laser fields not only can be applied to the ionization of atoms and molecules \cite{Joachain:2012,Liu:2014,Mima:2018} and the acceleration of charged particles \cite{Mangles:2004,Geddes:2004,Faure:2004} but also  provide
an alternative scheme to manipulate nuclear processes. For example, recent theoretical works have shown that the intense lasers can accelerate nuclear processes by resonance internal conversion \cite{Karpeshin:2006},  might increase $\alpha$ decay
 rates by modifying the Coulomb potential barrier \cite{Delion:2017,Qi:2019,Ghinescu:2020,Palffy:2020,Qi:2020} and
  excite the isomeric $^{229}{\rm Th}$ by laser-driven electron recollision \cite{Wang:2021}.

  Deuteron-triton (DT) fusion  in strong electromagnetic fields is also intriguing and has attracted recent attentions \cite{Queisser:2019,Lv:2019,Wang:2020,Liu:2021,Kohlfu:2021,Lv:2021,Qi:2021}, partly due to  its potential applications as a clear, effective, and sustainable energy in the future \cite{Atzeni:2004,Kikuchi:2011}.
   In the high-frequency regime of laser fields, the tunneling probability of DT fusion could be enhanced in light of the  Floquet scattering method \cite{Queisser:2019} and  Kramers approximation \cite{Lv:2019,Lv:2021}, respectively. In the low-frequency regime of laser fields, intense lasers are highly effective in transferring field energy to the DT system and therefore enhance the fusion probabilities based on quantum Volkov state approximation \cite{Wang:2020}. Moreover, it is found that  the tunneling penetrability can be enhanced dramatically in a wide range of laser parameters\cite{Liu:2021} or even by a time-dependent pulse-shaped vector potential \cite{Kohlfu:2021}.

  Based on Gamow tunneling picture \cite{Gamow:1928,Bosch:1992,Adelberger:1998}, the  DT fusion cross section can be  written as a product of geometrical cross section, the tunneling probability through the  Coulomb repulsive potential and astrophysical~$S$-factor. Previous studies  mainly focus on the enhancement effects
on tunneling probability  and  assume that the astrophysical~$S$-factor keeps identical to that in the absence of
strong electromagnetic fields\cite{Queisser:2019,Lv:2019,Wang:2020,Liu:2021,Kohlfu:2021,Lv:2021,Qi:2021}. Whether strong electromagnetic fields would change astrophysical~$S$-factor remains unclear.

Astrophysical~$S$-factor represents mainly the nuclear part of fusion cross section. Due to the lack of a full understanding of nuclear potential during fusion, the phenomenological complex potentials, also known as the "optical model" \cite{Dickhoff:2018}, are exploited to readily calculate fusion cross section within the framework of quantum scattering theory. A simple complex spherical square-well optical potential is widely applied to describe
the nuclear potentials of light nuclear fusion \cite{lixingzhong:2000,lixingzhong:2008,Singh:2019,Wu:2022} where an imaginary part of potential implies the decay of compound nucleus. This model only contains three parameters and can overcome the insufficiencies of Gamow tunneling formula: it shows that the tunneling and decay of compound nucleus are no longer  independent in light nuclear fusion process and need to be combined as a selective resonant tunneling \cite{Singh:2019}.

In this work, applying the simple optical potential model of complex spherical square-well to describe nuclear potential of DT fusion, we attempt to investigate nuclear fusion cross sections in presence of strong electromagnetic fields. Within the framework of Kramers-Henneberger (KH) transformation \cite{Henneberger:1968}, we have calculated  the total as well as the angular differential cross
sections with respect to varied parameters of the electromagnetic fields. The corresponding astrophysical $S$-factors are found to be enhanced by several times in amplitudes.
With the help of  Wentzel-Kramers-Brillouin (WKB) approximate wavefunctions \cite{Berry:1972}, the
shape-resonance tunneling mechanism of the above findings are uncovered and some implications are discussed.

This paper is organized as follows. In Sec. \uppercase\expandafter{\romannumeral2}, we present our theoretical model. In Sec. \uppercase\expandafter{\romannumeral3}, we present our main results and discussions. Sec. \uppercase\expandafter{\romannumeral4} is  conclusion.

\section{THEORETICAL MODEL}\label{method}

\subsection{KH transformation}

In the presence of electromagnetic fields, the relative motion of a spinless DT fusion system in the center-of-mass frame (CM) can be described by the time-dependent Schr\"{o}dinger equation
\begin{equation}
i\hbar\frac{\partial}{\partial t}\Psi(t,\boldsymbol{r})=\left(\frac{1}{2m}\left(\boldsymbol{p}-\frac{q_{\rm eff}}{c} \boldsymbol{A}(t,\boldsymbol{r})\right)^2+V(r)\right)\Psi(t,\boldsymbol{r}),
\end{equation}
where $m=m_{\rm d} m_{\rm t}/(m_{\rm d}+m_{\rm t})$ is reduced mass of DT. $m_{\rm d}$ and $m_{\rm t}$ are masses of deuteron and triton, respectively. $q_{\rm eff}=e(Z_{\rm d} A_{\rm t} -Z_{\rm t} A_{\rm d})/(A_{\rm d}+A_{\rm t})=e/5$ is an effective charge, where $Z_{\rm d}$ ($Z_{\rm t}$) and $A_{\rm d}$ ($A_{\rm t}$) are charge numbers and and mass numbers of deuteron (triton), respectively. $c$ is the speed of light in vacuum.
When the characteristic wave length of the field is much greater than the size of a typical nucleus (fm), the dipole approximation can be used so that the spatial dependency of the vector potential can be neglected, i.e., $\boldsymbol{A}(t,\boldsymbol{r})\approx\boldsymbol{A}(t)$.

A complex potential, also known as the "optical model" \cite{Dickhoff:2018}, can describe the scattering and absorption effects of particles. The superiority of absorptive nuclear force compared to the Coulomb repulsive leads to an absorptive potential well for DT fusion in the range of nuclear force. In this work, the potential $V(r)$ is considered to short-range complex spherical square potential well with a long-range Coulomb repulsive potential between two nuclei
\begin{equation}
V(r)=\left\{
\begin{array}{rcl}
V_r+{\rm i}V_i& & {r < r_n,}\\
\frac{ e^2}{4\pi \epsilon_0 r} & & {r > r_n,}\\
\end{array} \right.
\end{equation}
where $r_n=r_0(A_t^{1/3}+A_d^{1/3})$ is radius of nuclear well. Comparing with experimental benchmark cross section data in the absence of electromagnetic fields, the three optical parameters $V_r$, $V_i$ and $r_0$ can be calibrated. The optical parameters $V_r$ and $r_0$ of DT fusion are approximately 30 to 40 MeV and 1-2 fm, respectively \cite{Atzeni:2004}.

By applying the unitary KH transformation,
\begin{equation}
\Omega(t)={\rm exp}\left[\frac{i}{\hbar}\int^t_{-\infty}H_{\rm int}(\tau)d\tau\right]
\end{equation}
with
\begin{equation}
H_{\rm int}(\tau)=-\frac{q_{\rm eff}}{mc}\boldsymbol{A}(\tau)\cdot\boldsymbol{p}+\frac{q_{\rm eff}^2}{2mc^2}\boldsymbol{A}^2(\tau),
\end{equation}
the new wave function $\Phi\equiv\Omega(t)\Psi$ satisfies the following equation:
\begin{equation}\label{KH_TDSE}
i\hbar\frac{\partial}{\partial t}\Phi(t, \boldsymbol{r}_{\rm kh})=\left(\frac{ \boldsymbol{p}^2}{2m}+V(t,\boldsymbol{r}_{\rm kh})\right)\Phi(t,\boldsymbol{r}_{\rm kh}),
\end{equation}
where the time-dependent potential is found to be
\begin{equation}
V(t,\boldsymbol{r}_{\rm kh})=\left\{
\begin{array}{rcl}
V_r+{\rm i}V_i& & {r_{\rm kh}(t)<r_n,}\\
\frac{ e^2}{4\pi \epsilon_0 r_{\rm kh}(t)} & & {r_{\rm kh}(t)>r_n.}\\
\end{array} \right.
\end{equation}
The new coordinate operator $\boldsymbol{r}_{\rm kh}(t)$ is
\begin{equation}
\boldsymbol{r}_{\rm kh}(t)=\boldsymbol{r}-\boldsymbol{r}_e=\boldsymbol{r}-\frac{q_{\rm eff}}{mc }\int_{-\infty}^{t}\boldsymbol{A}(\tau)d\tau.
\end{equation}
where $\boldsymbol{r}$ is relative displacement vector of deuteron and triton and $\boldsymbol{r}_e$ is the
quiver displacement vector of a free nucleus in the electromagnetic fields.

Supposing that the external electromagnetic field is monochromatic and linearly polarized along the $z$ axis,
\begin{equation}
\boldsymbol{A}(t)=\hat{e}_z A_0 {\rm cos}(\omega t),
\end{equation}
then $r_{\rm kh}(t)=\sqrt{r^2-2r{\rm cos}(\theta) r_e {\rm sin}(\omega t) +(r_e {\rm sin}(\omega t))^2}$ with $r_e=e\sqrt{2c\mu_0I}/5m\omega^2$ can be obtained, where $\theta$ is the inclination angle between $\boldsymbol{p}$ and the polarization direction $z$ of the electromagnetic field, $\omega$ and $I$ are frequency and intensity of the electromagnetic field, respectively.
 It is useful to introduce the dimensionless quantity $n_{\rm d}=r_e/r_n=4.89\times10^{-6}\sqrt{I}/(\hbar\omega)^2$, where the
units of $I$ and $\hbar\omega$ are $ \rm W/cm^2$ and eV, respectively \cite{Delion:2017,Lv:2019,Lv:2021}.

The time-dependent potential $V(t,\boldsymbol{r}_{\rm kh})$ becomes axially deformed
and it can be expanded in a Fourier basis as follows,
\begin{equation}
  V(t,\boldsymbol{r}_{\rm kh})=\sum_{-\infty}^{\infty} V_n(\boldsymbol{r})e^{in\omega t},
\end{equation}
with expansion coefficient of
\begin{equation}
  V_n(\boldsymbol{r})=\frac{1}{T}\int_{0}^{T} V(t,\boldsymbol{r}_{\rm kh})e^{-in\omega t} \,dt,
\end{equation}
where $\omega=\frac{2\pi}{T}$ and $\boldsymbol{r}\equiv (r,\theta)$.

\subsection{Time-averaged potential}

The lifetime of the  particle is determined by the imaginary part of
the potential $V_i$ in the optical potential  model \cite{lixingzhong:2000}. When the lifetime  $\tau$ is much longer than
the field period, i.e., $\tau=h /V_i > 2\pi/\omega$, the fusion nucleus feels
an effective  time-averaged static potential. It is worthy noting that in this situation the laser period is  also much shorter than the characteristic collision time duration of the approaching nucleus in the combined Coulomb potential and electromagnetic fields, which is condition of Kramers approximation \cite{Lv:2019}. Thus,
 we choose $V_i/\hbar$  as the threshold of the laser
frequency beyond which the time-averaged static potential  is expected to be valid. Therefore, $V(t,\boldsymbol{r}_{\rm kh})$ approximates to be
\begin{equation}\label{V0}
  \begin{aligned}
    V(t,\boldsymbol{r}_{\rm kh})\approx V_0(r,\theta,n_d)&=\frac{1}{T}\int_{0}^{T} V(t, \boldsymbol{r}_{\rm kh})\,dt\\
    &=\frac{1}{T}\int_{0}^{T}\left[\Theta\left(1-\frac{r_{\rm kh}(t)}{r_n}\right)(V_r+iV_i)\right.\\
    &~~~~~~~~~~~~\left.+\Theta \left(\frac{r_{\rm kh}(t)}{r_n}-1\right)\frac{ e^2}{4\pi \epsilon_0 r_{\rm kh}(t)}\right]\,dt,
    \end{aligned}
\end{equation}
where $\Theta$ is the unit step function. The time-averaged potential $V_0$
 contains an angle-dependent  complex  nuclear potential  and the repulsive potential with a Coulomb tail. For convenience of following analytic treatment, we further approximate the nuclear potential by a complex square  well through a spatial average, i.e.,
 \begin{equation}\label{V0n}
 \overline V_0(r,\theta,n_d)=\left\{
\begin{array}{rcl}
\overline{V_r} (\theta,n_d)+i\overline{V_i} (\theta,n_d)& & {r<r_{n}(\theta,n_d),}\\
\frac{1}{T}\int_{0}^{T}\frac{ e^2}{4\pi \epsilon_0 r_{\rm kh}(t)} dt & & {r>r_{n}(\theta,n_d)},\\
\end{array} \right.
\end{equation}
where $\overline{V_r} (\theta,n_d)$, $\overline{V_i}(\theta,n_d)$, and $r_{n}(\theta,n_d)$ are a new set of effective  optical potential parameters of time-averaged nuclear potential.

\subsection{Complex phase shift}

From Eqs. \eqref{KH_TDSE} and \eqref{V0n}, the time independent Schr\"{o}dinger equation of DT fusion can be obtained as
$(-\hbar^2/2m)\nabla^2\psi(\boldsymbol{r})+(\overline V_0(r,\theta,n_d)-E)\psi(\boldsymbol{r})=0$,
where $E$ is relative kinetic energy of two nuclei in the center-of-mass frame. It is noted that the kinetic energy of incident projectile in the laboratory system,
$E_{\rm lab}=(m_p+m_t)/m_tE$. The potential  $\overline V_0(r,\theta,n_d)$ is  explicitly dependent  on angel $\theta$.  For this case, the stationary coupled
channels approach is applied to calculate the fusion cross section \cite{Ghinescu:2020}. Nevertheless, in this work,
 to avoid to solve complex coupled channel equations \cite{Massmann:1979}, for each angle $\theta$, the potential can be viewed as a central potential
 so that the stationary wave function can be written by $\psi(r,\theta,\phi)=Y_{l,m}(\theta,\phi)u_{l,\theta}(r)/r$.

For low-energy DT fusion ($E<1{\rm MeV}$), only the S wave ($l=0$) is considered. Then, the radial wave function satisfies the equations:
\begin{equation}\label{ur1}
\begin{split}
\frac{d^2u_{0,\theta}}{dr^2}+\frac{2m}{\hbar^2}\left(E-\overline V_{0,\theta}(r,n_d)\right)u_{0,\theta}(r)=0,\\
 u_{0,\theta}(r=0)=0.
\end{split}
\end{equation}

 The solution that satisfies the Eqs. \eqref{V0}, \eqref{V0n} and \eqref{ur1} can be written as \cite{Landau:1987}
\begin{equation}\label{ur2}
  u_{0,\theta}(r)=\left\{
\begin{aligned}
&B{\rm sin}(k_{n,\theta} r) ~r<r_{n}(\theta),\\
&D\left[{\rm F}_{0,\theta}(k,r){\rm cot}(\delta_{0,\theta})+{\rm G}_{0,\theta}(k,r)\right]~ r\rightarrow \infty,\\
\end{aligned}
\right.
\end{equation}
where $B$ and $D$ are constant coefficients, $k_{n,\theta}=\sqrt{2m(E-\overline{V_r} (\theta,n_d)-i\overline{V_i} (\theta,n_d)/\hbar^2}$ is the complex nuclear wave number, $\delta_{0,\theta}(E)$ is the complex phase shift of S wave, $k=\sqrt{2m E/\hbar^2}$ is the free particle wave number, ${\rm F}_0(kr,\eta)$ and ${\rm G}_0(kr,\eta)$ are regular and irregular Coulomb wave functions \cite{Abramowitz:1972}, respectively, where $\eta=1/k a_c$ is dimensionless Coulomb parameter, $a_c=4\pi\epsilon_0\hbar^2/m Z_d Z_t e^2$ is Coulomb unit length.

The phase shift $\delta_{0,\theta}(E)$ can be numerically obtained by solving radial Eq. \eqref{ur1} with restricted conditions Eq. \eqref{ur2} with the standard 4-5th
Runge-Kutta algorithm.
Then, the angle-dependent differential fusion cross section can be put in the form,
\begin{equation}\label{sigma_angle}
\sigma(\theta, E, n_d )=\frac{1}{4k^2}\left(1-|e^{2i\delta_{0,\theta}}|^2\right).
\end{equation}
The total cross section can be obtained by taking an integration  over the $4\pi$ solid angle:
\begin{equation}\label{sigma_total}
  \sigma_{\rm t}(n_d,E)=2\pi\int_{0}^{\pi} \sigma(\theta, n_d, E){\rm sin}\theta\,d\theta.
  \end{equation}

\section{NUMERICAL RESULTS AND DISCUSSION}\label{result1}

\subsection{Time-averaged potentials with varied field parameters}

In this work, we choose the optical potential parameters of DT nuclei as $V_{\rm r}$=-30.00 MeV, $V_{\rm i}$=-49.64 keV and $r_{\rm 0}=1.338$ fm \cite{Wu:2022}.
 \begin{figure}[!b]
\includegraphics[width=0.50\textwidth]{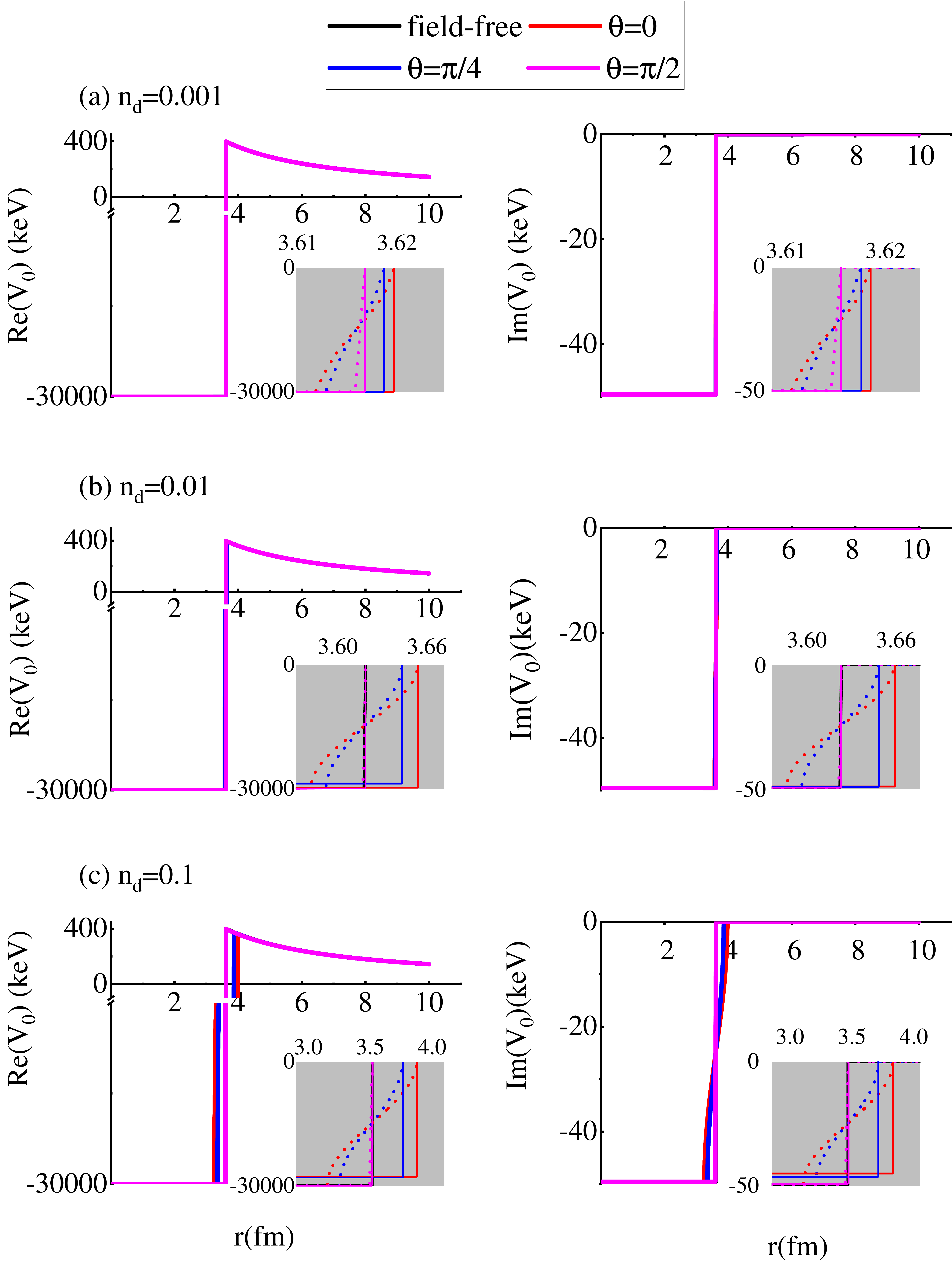}
\vspace{-5mm}
\centering
\caption{(Color online) The time-averaged potential for
different inclination angles $\theta$ with respect to varied $n_d$: the gray subplots represent the zoom of nuclear potential, where the dotted lines are time-averaged nuclear well and the solid lines are approximate ones.}
\label{fig:1}
\end{figure}

The time-averaged optical potential $V_0(r,\theta,n_d)$ has rotational symmetry with respect to
$z$ axis and is  independent on the azimuth angle $\varphi$. We plot $V_0(r,\theta,n_d)$ for
 varied inclination angles of $\theta=0,\pi/4$ and $\pi/2$, and $n_d=0.001, 0.01$ and $0.1$ in Fig. \ref{fig:1}, respectively. Comparing with field-free case, Fig. \ref{fig:1} shows that $V_0(r,\theta,n_d)$  consists of a slightly deformed nuclear square well and
 a repulsive potential with a Coulomb tail  for the fixed $\theta$ and $n_d$.

In order to clearly show the deformation of potential, the zooms of the nuclear potential are displayed in the subplots of Fig. \ref{fig:1}, where the dotted lines are time-averaged nuclear potential while the solid lines are the approximate square nuclear well. For a fixed $n_d$, as  shown in subplot of Fig. \ref{fig:1},  the $\theta=0$ corresponds to the largest deformation of $V_0(r,\theta,n_d)$ while the smallest deformation corresponds to $\theta=\pi/2$. We also can see from subplot of Fig. \ref{fig:1} that, for a fixed $\theta$, the deformation of $V_0(r,\theta,n_d)$ becomes larger with increasing the $n_d$.

The approximate square nuclear well can qualitatively reflects the field induced deformation  of nuclear potential. Comparing with the field-free case, the new square nuclear well (both real and imaginary part) becomes shallower and wider as shown in subplot of Fig. \ref{fig:1}, which can be expresses by a new set of optical potential parameters. These effective  optical potential parameters ($\overline{V_r} (\theta,n_d)$, $\overline{V_i}$, $(\theta,n_d)$, and $r_{n}(\theta,n_d)$) depends on the inclination angle $\theta$ as well as the dimensionless quantify $n_d$.  As will shown below, the slight changes on the optical potential parameters  will cause a huge change in the fusion cross section as well as corresponding $S$ factor due to resonant tunneling effect.

\subsection{The angle-dependent differential cross section of DT fusion}

\begin{figure}[!t]
\includegraphics[width=\linewidth]{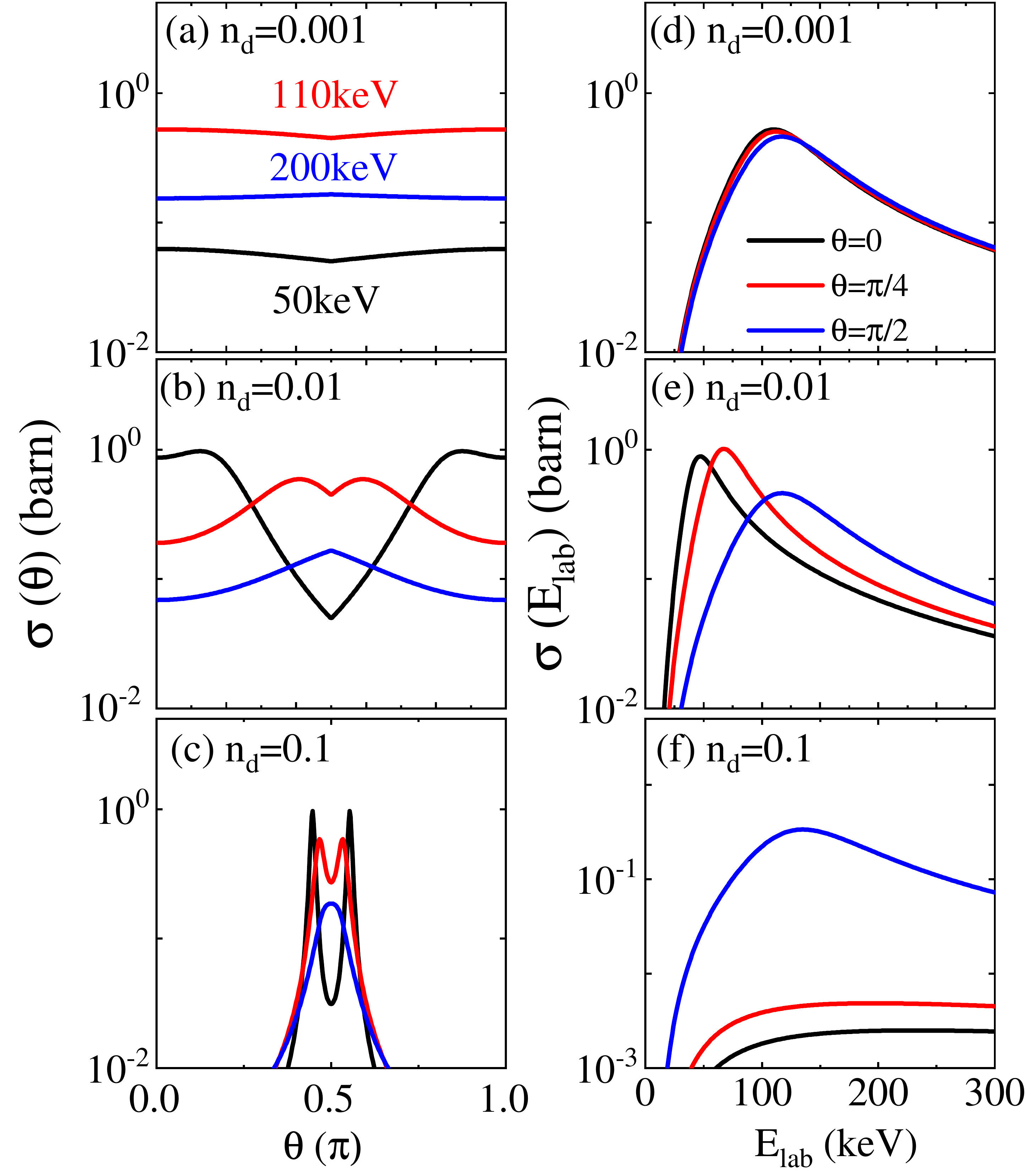}
\vspace{-5mm}
\centering
\caption{(Color online) The angle-dependent differential fusion cross section for $n_{\rm d}=0.001,0.01,0.1$ as a function of : (a)-(c) inclination angle $\theta$ for fixed energies $E_{\rm lab}=50,110,200$ keV; (d)-(f) energy $E_{\rm lab}$ for fixed angle $\theta=0,\pi/4,\pi/2$.}
\label{fig:2}
\end{figure}

We have calculated the angle-dependent differential cross section $\sigma(\theta, E_{\rm lab}, n_d )$ of DT fusion according to Eq. \eqref{sigma_angle} and shown the  results in Fig. \ref{fig:2}.

 Figs. \ref{fig:2} (a)-(c) display $\sigma(\theta, E_{\rm lab}, n_d )$ as a function of inclination angle $\theta$ with respect to  the varied incident particle energies $E_{\rm lab}$ for
 $n_d=0.001,0.01$ and 0.1, respectively.  The $\sigma(\theta, E_{\rm lab}, n_d )$ exhibits a symmetry with respect to $\theta=\pi/2$. It shows some resonance peaks that shift from zero angle to $\pi/2$ with increasing the parameter $n_d$.
  For $n_d=0.001$, the $\sigma(\theta, E_{\rm lab}, n_d )$  changes very slowly with increasing $\theta$ for  a fixed incident energy of $E_{\rm lab}$.

 However, for $n_d$=0.01 and 0.1, we can see from Figs. \ref{fig:2} (b)-(c) that the cross sections $\sigma(\theta, E_{\rm lab}, n_d )$ sensitively depends on the inclination angle $\theta$ and exhibits some interesting resonance structures.
 The $\sigma(\theta, E_{\rm lab}, n_d )$ appears double-peak structure and the angle corresponding to its peak has shifted for different energy $E_{\rm lab}$. For example, for $E_{\rm lab}=200$ keV, the $\sigma(\theta, E, n_d )$ can reach maximum at about $\theta=\pi/2$, i.e., the direction perpendicular to the electromagnetic field polarization direction, while for $E_{\rm lab}=50$ keV the peak turns to locate at about $\theta=\pi/5$ as shown in Fig. \ref{fig:2}(b).  Moveover, Fig. \ref{fig:2} (c) shows that the cross section $\sigma(\theta, E_{\rm lab}, n_d )$ has a larger value near $\theta=\pi/2$ and significantly decreases when the inclination angle is away  from $\theta=\pi/2$ . Note that, the peak of $\sigma(\theta, E_{\rm lab}, n_d )$ in the direction perpendicular to the field polarization (i.e. $\theta=\pi/2$) is quite interesting. Ref. \cite{Delion:2017} and \cite{Lv:2019} also find that at $\theta=\pi/2$, the angle-dependent tunneling probability can reach local maxima in the Karamers approximation. Differently, in our situations, tunneling probability is almost independent on inclination angle. The above  results can be understood by the shape-resonance mechanism as will discussed in the following sections.

 We also plot the $\sigma(\theta, E_{\rm lab}, n_d )$ as a function of energy $E_{\rm lab}$ for varied values of $n_d$  and three  angles of $\theta=0,\pi/4,\pi/2$ in the Figs. \ref{fig:2} (d)-(f), respectively.
In Fig. \ref{fig:2} (d), the curves for varied  angles are almost identical.
However, with increasing $n_d$ to $0.01$ and $0.1$, the deviations between  the curves corresponding to   different angles increase significantly. Moreover, the peaks of cross section shift from low-energy region to high-energy region with increasing $\theta$ for $n_d=0.01$ in the Fig. \ref{fig:2}(e). Fig. \ref{fig:2}(f)
shows that compared with that of $\pi/2$, the cross sections of $\theta=0,\pi/4$ can be significantly reduced.
\begin{figure}[!t]
\includegraphics[width=0.50\textwidth]{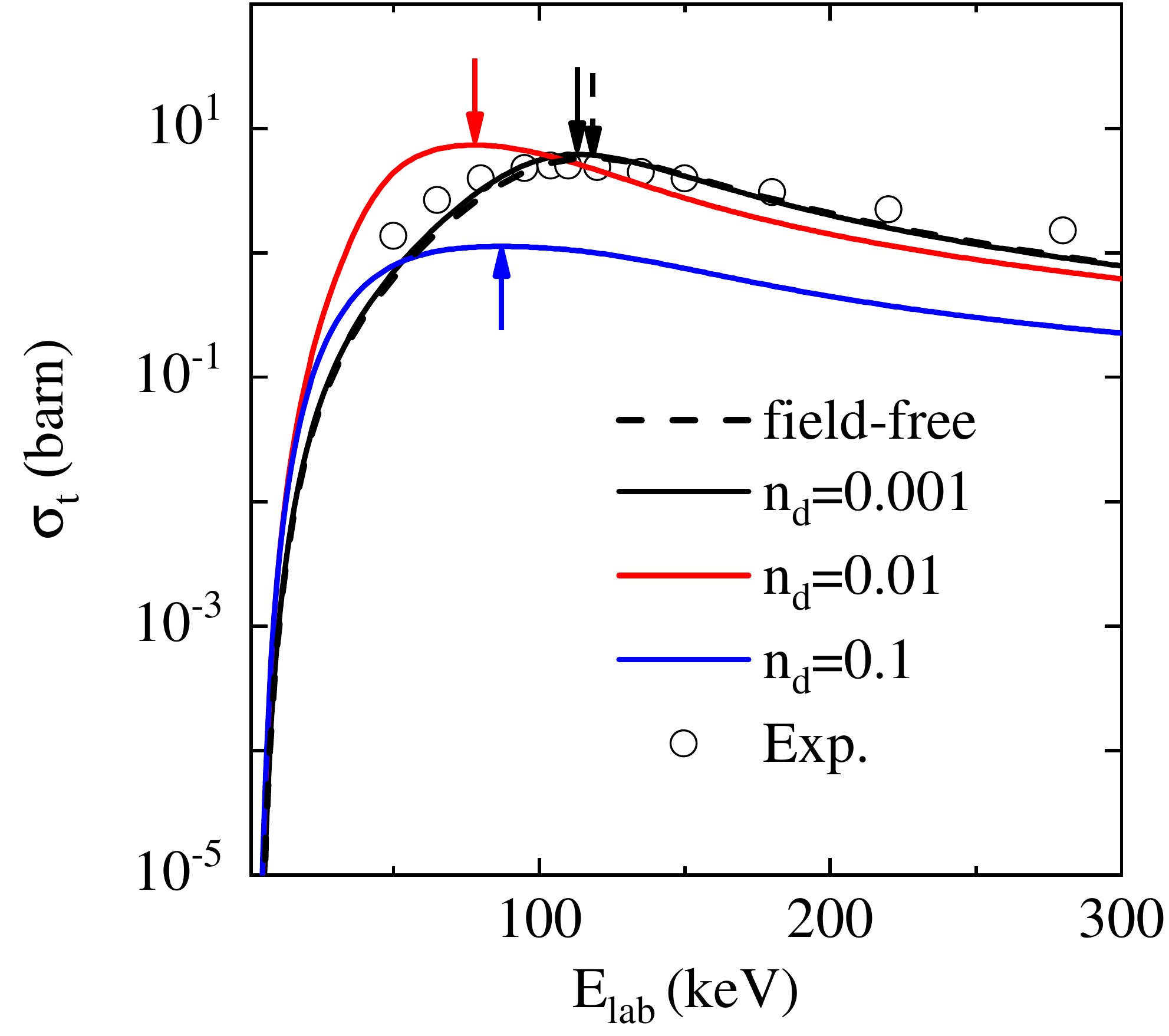}
\vspace{-5mm}
\centering
\caption{(Color online) Total DT fusion cross section $\sigma_{\rm t}(n_d,E_{\rm lab})$ versus collision energy $E_{\rm lab}$ for $n_d=0.001, 0.01$ and 0.1. The dotted line and the hollow circles represent cross section and the experimental data \cite{lixingzhong:2008} in absence of the electromagnetic field, respectively. The arrows represent the positions of the peak.}
\label{fig:3}
\end{figure}

\subsection{Total DT fusion cross section}

 According to Eq. \eqref{sigma_total}, we calculate total  DT fusion cross sections $\sigma_t(n_d,E_{\rm lab})$ as a function of collision energy $E_{\rm lab}$ for $n_d=0.001, 0.01$ and 0.1 in the Fig. \ref{fig:3}, respectively.
 In the absence of field, our calculations are consistent with the experimental data  \cite{lixingzhong:2008}.
 Comparing with that of field-free case, the total  fusion cross sections  are enhanced in low-energy region while descrease in high-energy region. The peaks (see the arrows in Fig. \ref{fig:3}) corresponding to the  maximum cross sections show an apparent shift from the well known value of 110keV to 78keV for $n_d=0.01$.
 Most interestingly, the peak value of $\sigma_{\rm t}(n_d,E_{\rm lab})$ increases first and then decreases with the increase of $n_d$, which implies that
 there exists an optimal $n_d$ to significantly enhance the $\sigma_{\rm t}(n_d,E_{\rm lab})$ in low-energy region.
\subsection{Astrophysical $S$-factor}

According to the Gamow tunneling picture \cite{Gamow:1928,Bosch:1992,Adelberger:1998}, astrophysical $S$-factor in strong electromagnetic field can be defined as
\begin{figure}[!t]
\includegraphics[width=0.50\textwidth]{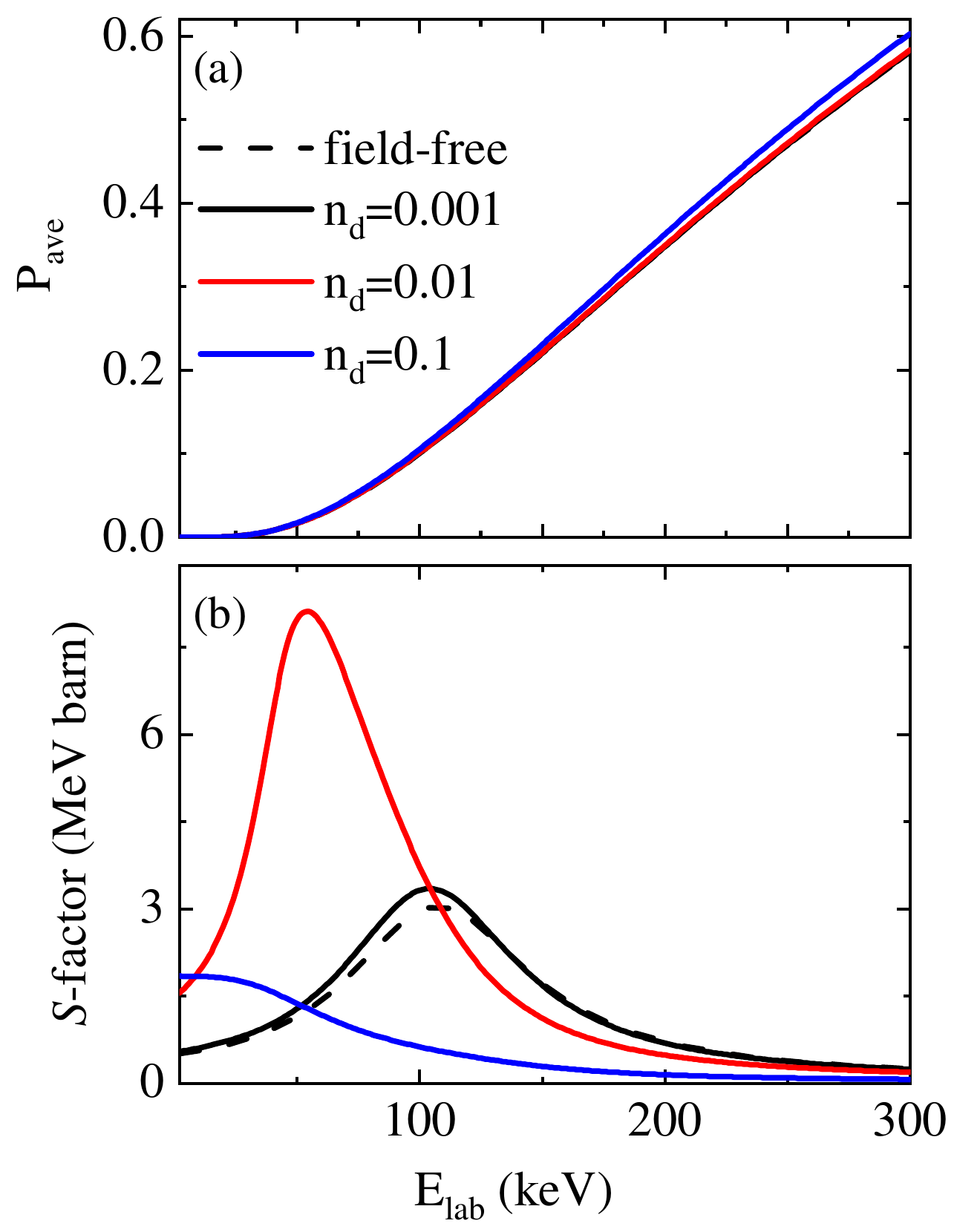}
\vspace{-5mm}
\centering
\caption{(Color online) (a) Average penetrability probability $P_{\rm ave}$  versus collision energy $E_{\rm lab}$ for $n_d=0.001,0.01$ and 0.1. (b) Corresponding $S$ factor. The dotted lines represent results in absence of the electromagnetic field. }
\label{fig:4}
\end{figure}
\begin{equation}\label{SE}
S(E,n_d)\equiv\frac{\sigma_t E}{P_{\rm ave}(E, n_d)},
\end{equation}
where $P_{\rm ave}(E, n_d)$ is the angle-averaged penetrability probability through the Coulomb repulsive barrier  modified by the strong electromagnetic field \cite{Lv:2019,Lv:2021}.  Fig. \ref{fig:4} (a) shows that penetrability probabilities are only slightly increased  with increasing the scaled parameter  $n_d$ from $0$ to $0.1$. While, we find that astrophysical $S$-factor has changed dramatically.
For instance,  as shown in Fig. \ref{fig:4} (b), compared with that of field-free case, the peak of $S$-factor for $n_d=0.01$ moves to the low-energy region and its value can be enhanced by several times in amplitude.

\subsection{Shape-resonance tunneling with WKB description}
\begin{figure}[!t]
\includegraphics[width=0.5\textwidth]{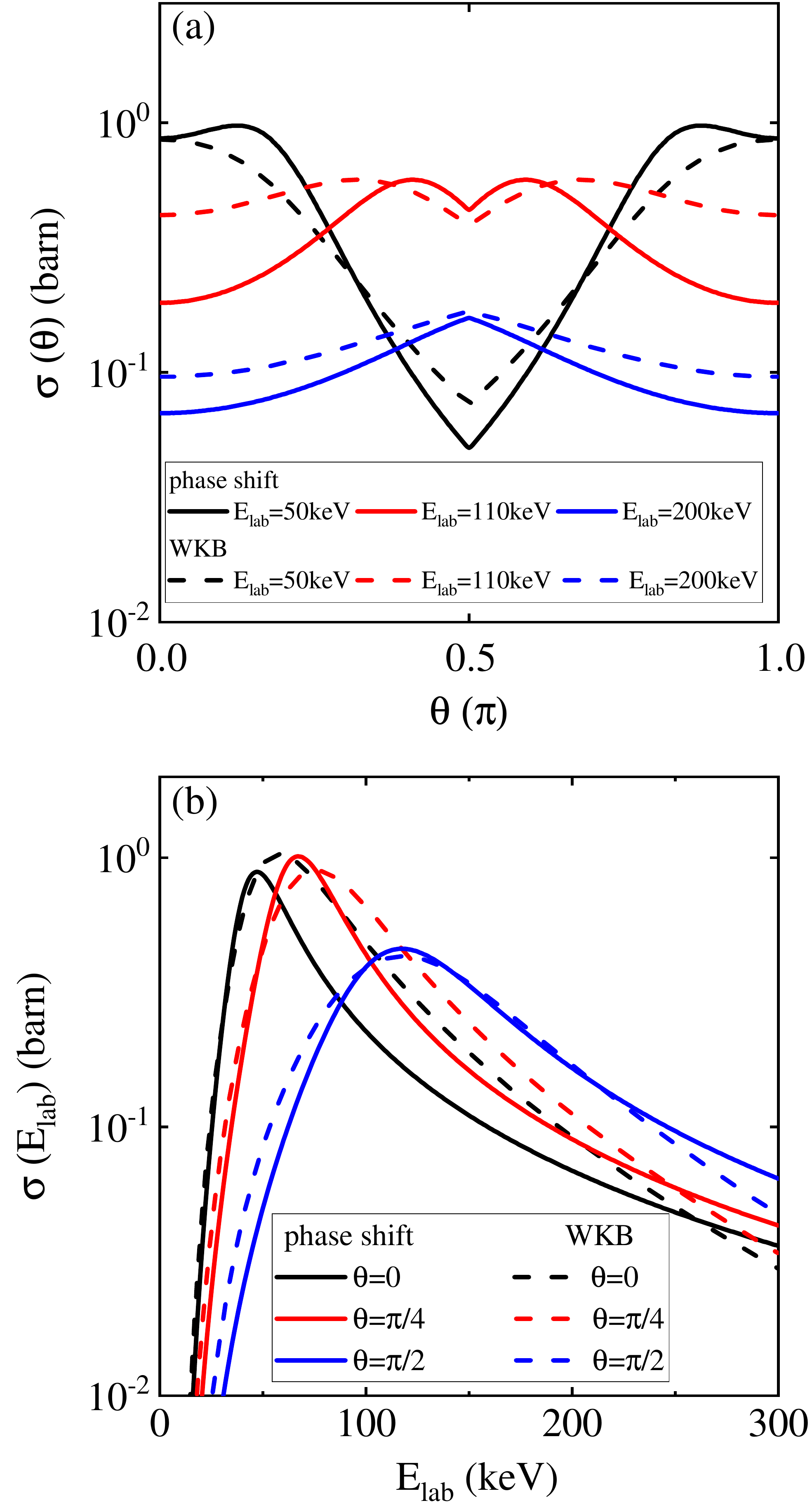}
\vspace{-5mm}
\centering
\caption{(Color online) Same as Fig. \ref{fig:2} but $n_d=0.01$: the phase shift results (solid lines) and the WKB results (dotted lines).}
\label{fig:5}
\end{figure}
In this section, we attempt to use the explicit WKB wavefunctions \cite{Berry:1972} to understand and explain our the above findings. With help of WKB description, the solution of radial Eq. \eqref{ur1} for each angle $\theta$ can be approximated as
\begin{equation}\label{ur3}
  u_{0,\theta}(r)=\left\{
\begin{aligned}
&A_1{\rm sin}(k_{n,\theta} r) ~r<r_{n,\theta},\\
&\sqrt{k/k_{2,\theta}}\left(B_1{\rm exp}\left(\int_{r_{n,\theta}}^{r}-k_{2,\theta}(r^{\prime})dr^{\prime}\right)\right.\\
&\left.+C_1{\rm exp}\left(\int_{r_{n,\theta}}^{r}k_{2,\theta}(r^{\prime})dr^{\prime}\right)\right)~ r_{n,\theta}<r<r_{\rm in,\theta},\\
&\sqrt{k/k_{1,\theta}}\left(D_1{\rm exp}\left(-i\left(\int_r^{r_{\rm in}}k_{1,\theta}(r^{\prime})dr^{\prime}+\frac{\pi}{4}\right)\right)\right.\\
&\left.+E_1{\rm exp}\left(i\left(\int_r^{r_{\rm in}}k_{1,\theta}(r^{\prime})dr^{\prime}+\frac{\pi}{4}\right)\right)\right)~ r>r_{\rm in,\theta},\\
\end{aligned}
\right.
\end{equation}
where $r_{\rm in}$ is classic turning point,
the local wave number is given by $k_{2,\theta}=k\sqrt{\overline V_{0,\theta}(r,n_d)/E-1}$ while $k_{1,\theta}=k\sqrt{1-\overline V_{0,\theta}(r,n_d)/E}$.
\begin{figure}[!t]
\includegraphics[width=0.5\textwidth]{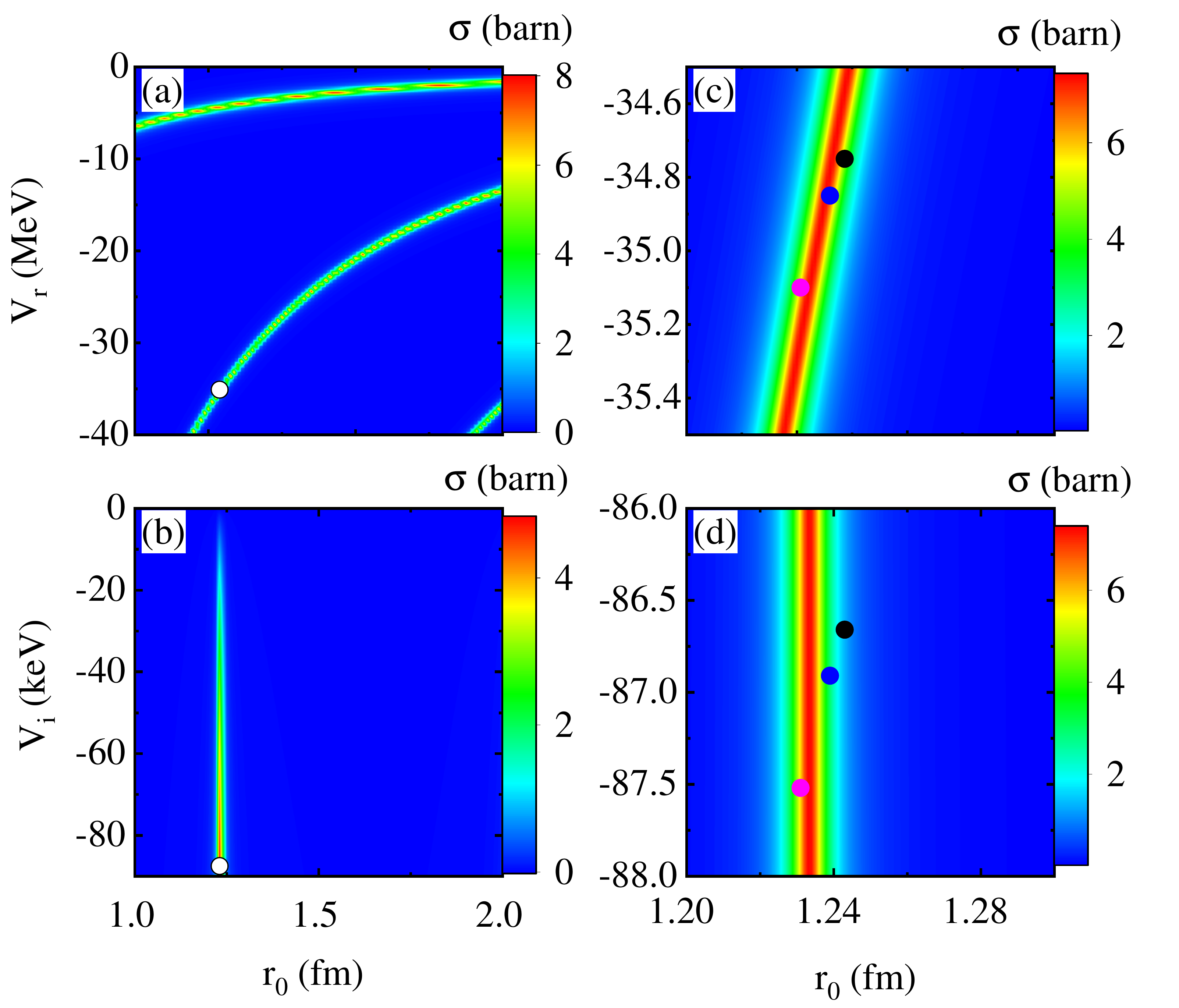}
\vspace{-5mm}
\centering
\caption{(Color online) Contour plots of fusion cross section based on WKB approximation for DT fusion reaction as a function
of  parameters of nuclear potential well at fixed energy $E_{\rm lab}=110$ keV, imaginary part $V_i=-87.52$ keV of (a) and (c), and real part $V_r=-35.10$ MeV of (b) and (d). The hollow circles represent the optical parameters in absence of the electromagnetic fields. The solid points are effective optical potential parameters of $n_d=0.01$ for three inclination angles $\theta$: $\theta=0$ (black points), $\theta=\pi/4$ (blue points) and $\theta=\pi/2$ (pink points). }
\label{fig:6}
\end{figure}
Note that the relation of coefficients $A_1$, $B_1$ and $C_1$ can be obtained  by the continuity conditions of wave function and its first derivative at $r=r_{n,\theta}$.
 The connection formulas of the coefficients between the
classical permitted and quantum tunneling regions can be written in following explicit form,
\begin{equation}
D_1=\frac{iB^{\prime}_1-2C^{\prime}_1}{2i},E_1=\frac{iB^{\prime}_1+2C^{\prime}_1}{2i}
\end{equation}
with
\begin{equation}
B^{\prime}_1=B_1 {\rm exp}\left(\int_{r_{\rm n,\theta}}^{r_{\rm in,\theta}}-k_{2,\theta}(r^{\prime})dr^{\prime}\right),
C^{\prime}_1=C_1 {\rm exp}\left(\int_{r_{\rm n,\theta}}^{r_{\rm in,\theta}}k_{2,\theta}(r^{\prime})dr^{\prime}\right).
\end{equation}

Then, the angle-dependent differential fusion cross section is given by
\begin{align}\label{WKB_sigma}
\sigma_{\rm wkb}(\theta, E, n_d)&=\frac{1}{4 k^2}\left(1-\left|\frac{E_1}{D_1}\right|^2\right)\notag\\
&=\frac{\hbar^2}{m E}\frac{-{\rm Im}(B_1C_1)}{4|C_1|^2-4{\rm Im}(B_1C_1) P+|B_1|^2P^2}P
\end{align}
where $P(n_d,\theta, E )={\rm exp}\left(\int_{r_{\rm n,\theta}}^{r_{\rm in,\theta}}-2k_{2,\theta}(r^{\prime})dr^{\prime}\right)$ is
the  the tunneling probability through the barrier.  The analytical Eq. \eqref{WKB_sigma} based on WKB clearly shows that the tunneling and decay of compound nucleus can no longer be independent in DT fusion process and need to be combined as a selective resonant tunneling \cite{lixingzhong:2000, Singh:2019}.
Note that, due to the sensitivity of  fusion cross section on the continuity conditions of wave function form \cite{Wu:2022}, we need recalibrate optical parameters of DT fusion based on WKB approximation to compare with field-free experimental cross sections.
\begin{figure}[!t]
\includegraphics[width=0.5\textwidth]{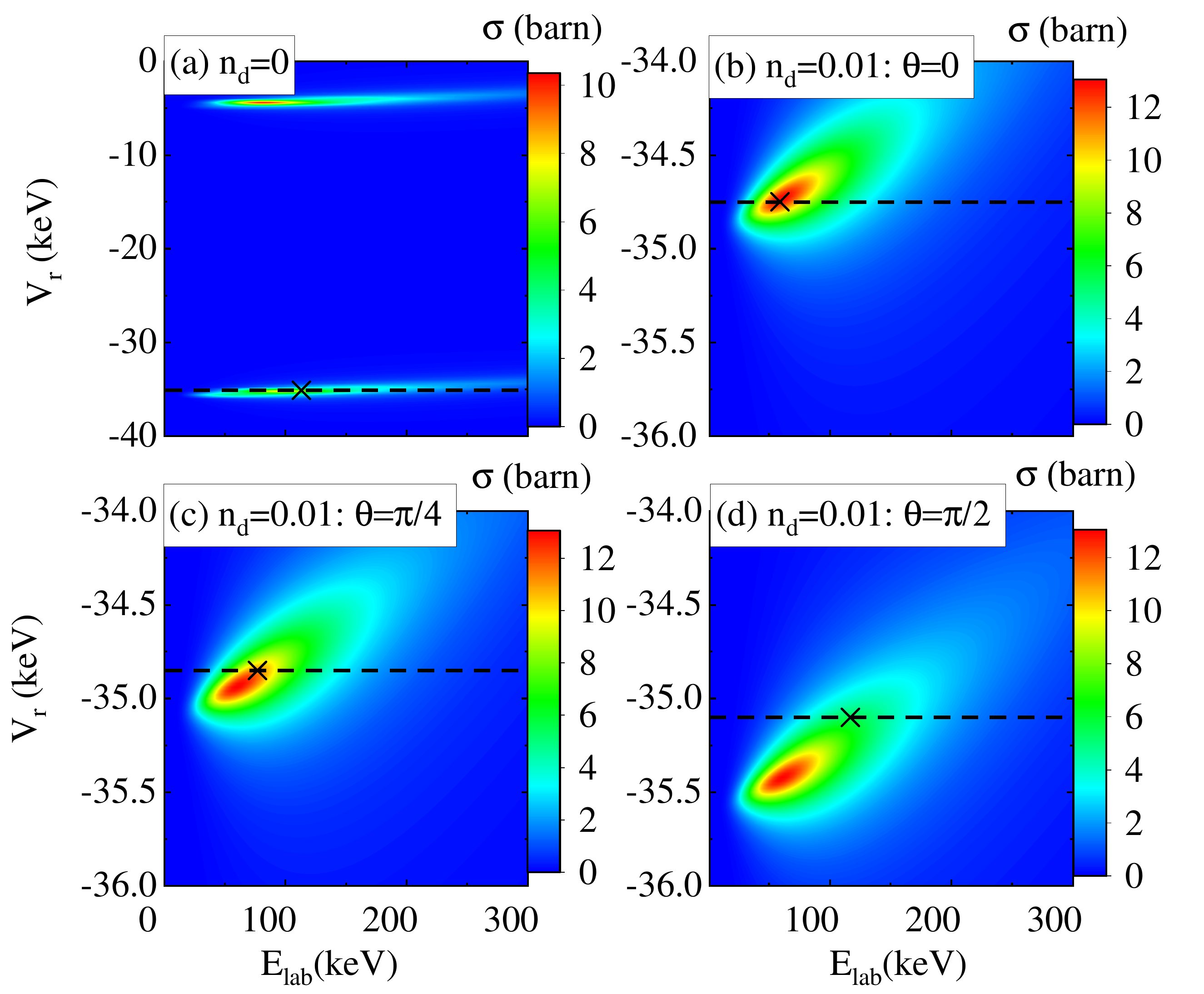}
\vspace{-5mm}
\centering
\caption{(Color online) The phase diagrams of cross section of DT fusion based on the WKB approximation method as a function of the real part $V_r$ and
 collision energy $E_{\rm lab}$ for different imaginary parts $V_i$ and radius parameters $r_0$.
  The fixed two parameters are the imaginary parts and radius of optical potential parameters of (a) $n_d=0$, (b) $n_d=0.01,\theta=0$, (c) $n_d=0.01, \theta=\pi/4$, and (d) $n_d=0.01, \theta=\pi/2$, respectively. The horizontal dashed lines represent the corresponding the real part of optical potential parameters. The black crosses are the positions of peak of the cross section along the horizontal dashed lines. }
\label{fig:7}
\end{figure}

We focus on case of $n_d=0.01$ as an example to make our analysis.
 The results (dotted lines) of the angle-dependent differential cross sections   are displayed in Fig. \ref{fig:5}. For comparison, we also show the results (solid lines) from the  phase shift method there. One can see that the trend of the curves from two methods is roughly the same, which suggests that the WKB approximation method is valid to be used  to analyze our results.

We then plot the phase diagrams of total cross section for DT fusion with WKB approximation method as a function of varied parameters of nuclear potential at the fixed energy $E_{\rm lab}=110$ keV in Fig. (6).
  The hollow circles represent the optical potential parameters in absence of the electromagnetic fields.  Fig \ref{fig:6} (a) demonstrates several narrow  resonance belts and the hollow circle is exactly in the resonance regime. Meanwhile, Fig. \ref{fig:6} (b) shows a narrow vertical belt and the fusion cross section monotonically decrease with increasing $V_i$. Note that, the resonance belts are induced by the  sensitive dependence of the cross sections on the nuclear well's shape, is a kind of shape resonance in quantum mechanics \cite{Combes:1987}.

 The effective optical potential parameters (i.e.,$\overline{V_r}(\theta,n_d)$, $\overline{V_i}(\theta,n_d)$, and $r_{n}(\theta,n_d)$ ) of  time-average optical potential are slight different from that of the field-free case, which depends on the inclination angel $\theta$ and dimensionless quantify $n_d$. We enlarge the corresponding parts of the phase diagrams of Figs. \ref{fig:6} (a)-(b) and mark the effective optical potential  parameters of $n_d=0.01$ for three inclination angles in Figs. \ref{fig:6} (c)-(d): $\theta=0$ (black points), $\theta=\pi/4$ (blue points) and $\theta=\pi/2$ (pink points), respectively. One can see from Figs. \ref{fig:6} (c)-(b) that, with increasing $\theta$, the scattered points will pass
 through the shape resonance belts. This might lead to appearance of the  double-peak structure of the angle-dependent
 differential cross section in Figs. \ref{fig:2} (b) and \ref{fig:5} (a).

 In order to deeply understand the peak shift of collision energy dependent total as well as the angular differential cross sections in Figs \ref{fig:2}, \ref{fig:3} and \ref{fig:5}, we also plot the phase diagrams of cross sections as a function of the real part $V_r$ and
 collision energy $E_{\rm lab}$ for different imaginary parts $V_i$ and radius parameters $r_0$ in Fig. \ref{fig:7}. The horizontal dashed lines represent the corresponding the real part of optical potential parameters. The black crosses are the positions of peak of the cross section along the horizontal dashed lines.

 The phase diagrams of Fig. \ref{fig:7} (a) show clearly two horizontal belts in the absence of the electromagnetic fields. Comparing with field-free case, from the Figs. \ref{fig:7} (b)-(d), we can see that the slight changes on nuclear potentials (or corresponding effective optical potential parameters)  due to the presence of  electromagnetic fields will move the position of resonant belts.
Then, with increasing $\theta$, the collision energy (see black crosses in the Figs. \ref{fig:7} (b)-(d)) corresponding to maximum  cross section shows a shift from low energy to high energy. This might
lead to the apparent shift of peak for total as well as angular differential cross section in Figs \ref{fig:2}, \ref{fig:3} and \ref{fig:5}

\section{Conclusions}

In conclusion, with applying a simple complex spherical square-well optical model and exploiting Kramers approximation, we have investigated DT fusion cross sections  in the presence of  strong high-frequency electromagnetic fields.
We find that the  peaks of  total cross sections as well as angular differential cross sections show an apparent shift due to  the mechanism of shape-resonance tunneling. The corresponding astrophysical $S$-factor can be enhanced by several times in amplitude, which imply that the strong electromagnetic field can significantly affect the nuclear part of fusion cross section. These results also suggest that the demanding Lawson criterion \cite{Lawson:1957} of controlled DT fusion may be relaxed in the presence of  strong  electromagnetic fields.
 In this work, we only consider the first order (or static term ) of  Fourier expansions in KH transformation. Higer-order terms  of the Fourier expansion should be further investigated especially for the cases of   low or medium frequencies.  Moveover, the complex spherical square-well is the simplest optical optical to describe the nuclear potential.  More realistic optical
 potential with rigid core such as Woods-Saxon nuclear potential \cite{Newton:2004} and nuclear spin effects \cite{Hupin:2019} in the strong electromagnetic fields are worthy of future's study. On the other hand,  the optical potential used in the present work is a phenomenological model based on mean field approximation. Within the framework
 of quantum chromodynamics (QCD), the \emph{ab initio} many-body calculations \cite{Petr:2012} of DT fusion in the strong electromagnetic fields are of great interest and challenging topics worthy of  further consideration.


\begin{acknowledgments}

\noindent
We thank Dr. Wenjuan Lv for some useful discussions. This work was supported by funding from NSFC No.\ 11775030 and NSAF No. U\ 1930403.

\end{acknowledgments}

\end{document}